\documentclass{article}
\usepackage{ijcai25}

\usepackage{makecell}
\usepackage{multirow}

\usepackage{times}
\usepackage{soul}
\usepackage{url}
\usepackage[hidelinks]{hyperref}
\usepackage[utf8]{inputenc}
\usepackage[small]{caption}
\usepackage{graphicx}
\usepackage{amsmath}
\usepackage{amsthm}
\usepackage{booktabs}
\usepackage{algorithm}
\usepackage{algorithmic}
\usepackage[switch]{lineno}

\title{Q-Detection: A Quantum-Classical Hybrid Poisoning Attack Detection Method}

\author{
Haoqi He\and
Xiaokai Lin\and
Jiancai Chen\And
Yan Xiao$^*$\\
\affiliations
School of Cyber Science and Technology, Shenzhen Campus of Sun Yat-sen University\\
\emails
\{hehq23, linxk5, chenjc26\}@mail2.sysu.edu.cn,
xiaoy367@mail.sysu.edu.cn
}

\begin{document}

\maketitle
\let\thefootnote\relax
\footnotetext{$^*$ Corresponding author. }

\begin{abstract}
    Data poisoning attacks pose significant threats to machine learning models by introducing malicious data into the training process, thereby degrading model performance or manipulating predictions. Detecting and sifting out poisoned data is an important method to prevent data poisoning attacks. Limited by classical computation frameworks, upcoming larger-scale and more complex datasets may pose difficulties for detection. We introduce the unique speedup of quantum computing for the first time in the task of detecting data poisoning. We present Q-Detection, a quantum-classical hybrid defense method for detecting poisoning attacks. Q-Detection also introduces the Q-WAN, which is optimized using quantum computing devices. Experimental results using multiple quantum simulation libraries show that Q-Detection effectively defends against label manipulation and backdoor attacks. The metrics demonstrate that Q-Detection consistently outperforms the baseline methods and is comparable to the state-of-the-art. Theoretical analysis shows that Q-Detection is expected to achieve more than a 20\% speedup using quantum computing power. 

\end{abstract}

\section{Introduction}

In modern image classification tasks, training classification models typically requires a large amount of data, and this data often comes from external sources such as web scraping, data marketplaces, or crowdsourcing platforms. While this method of data acquisition brings rich data resources, it also introduces potential security risks: malicious data providers can manipulate data to degrade model performance or interfere with its prediction behavior as ~\cite{datapoison1} said.

Data poisoning attacks are a common form of data contamination. These attacks alter the training data (features, labels, or both) to degrade model performance or exert control over it. Numerous studies as ~\cite{simpleButHeavy1,simpleButHeavy2} have found that even simple noise or malicious data points can significantly impact deep learning models. When it comes to critical domains such as autonomous driving or medical diagnosis, such attacks can severely mislead image classification models, leading to serious safety hazards.

Data poisoning attacks mainly include the following forms:
\begin{itemize}
    \item \textbf{Label-only attacks}: Only modify the labels without changing the features, such as Target Label-Flipping Attacks ~\cite{jha2023label}.
    \item \textbf{Feature-only attacks}: Only manipulate the features without changing the labels, such as Narcissus Backdoor Attacks ~\cite{zeng2023narcissus}.
    \item \textbf{Label-Feature attacks}: Modify both labels and features simultaneously, such as BadNets attacks ~\cite{gu2019badnets}.
\end{itemize}

To counter these attacks, researchers have proposed a series of defense methods, including data augmentation, adversarial training, anomaly detection, etc~\cite{cleandatasetneed1,cleandatasetneed2,cleandatasetneed3}. 

However, these methods usually rely on \textbf{clean benchmark datasets} for training or validation. Data poisoning detection methods on large datasets may encounter large time complexity. In recent years, quantum technology has shown potential for acceleration in complex optimization tasks ~\cite{hhq1}. By using quantum methods in the detection process, it is possible to obtain dual benefits in terms of efficiency and performance against poisoning. To our knowledge, the application of quantum computing in the defense or detection of data poisoning attacks is very rare ~\cite{quantumapply1,quantumapply2}.
Therefore, how to extract a clean subset by quantum method to avoid interference from poisoned samples to model performance has become a critical issue.

In this work, we explore the intersection of quantum computing and data poisoning detection, designing a quantum-classical hybrid poisoning attack detection method to filter clean training datasets for image classification tasks. This method ensures that models trained on the clean subset achieve leading accuracy and are expected to benefit from quantum computational speedups in the future.

We propose \textbf{Q-Detection}, a quantum-classical hybrid poisoning attack detection method. Q-Detection is based on the fact that the distribution between poisoned and clean data shifts, leading to higher losses during cross-validation. It transforms the poisoning detection problem into a bilevel optimization problem, maximizing the loss of the model trained on one subset when evaluated on another subset. Q-Detection ensures model training occurs in a clean environment by eliminating poisoned samples, resulting in deep learning models with higher accuracy.

Q-Detection establishes a \textbf{Quantum Weight-Assigning Network (Q-WAN)} that can be trained using quantum devices. For the first time, it transforms the selection problem into a QUBO optimization problem for the Q-WAN, enabling training with different quantum devices (such as quantum annealers, coherent optical quantum computers, superconducting gate model quantum computers). This provides a new direction for integrating quantum computing into deep learning in the future. 

When filtering clean data, the Q-WAN will train alongside a domain model. It is utilized the domain model's prediction errors during training. By inputting different data to generate different prediction errors, the Q-WAN's parameters are dynamically adjusted, enabling different weights to be assigned to different data samples. These weights are then used to filter and identify clean samples, suppressing poisoned samples. 

Our contributions are as follows:
\begin{itemize}
    \item Q-Detection enhances the integration of quantum computing and security fields. To our knowledge, it is the first quantum-classical hybrid method to transform the selection problem into a bilevel QUBO optimization problem, introducing quantum computational power. It is the first quantum method to complete data poisoning detection and clean dataset selection tasks.

    \item Q-Detection proposes a QUBO-based bilevel formulation method that serves as a unifying backbone, enabling training of the Q-WAN using different quantum computing technologies.
    
    \item Q-Detection achieves SOTA and consistently outperforms baseline methods across three different poisoning attacks and various poisoning ratios, demonstrating its advanced capabilities. Furthermore, the selected subset used to train new models achieves accuracy surpassing baseline methods on both global and specific label data. 

    \item Q-Detection is enhanced by the progress of the times: our experiments reveal that the hidden layer size of the Q-WAN, linearly correlated with the number of qubits, determines Q-Detection's selection capability. Thus, it can be anticipated that with breakthroughs in quantum computing technology and the gradual increase in qubits, Q-Detection will have a broader prospect in the future.
\end{itemize}

The demonstration experiment code can be found at https://github.com/cats1520cakes/Q-Detection-A-Quantum-Classical-Hybrid-Poisoning-Attack-Detection-Method.

\section{Preliminaries}

This section briefly reviews representative instances in Label-only, Feature-only, and Label-Feature attacks (such as Random Label-Flipping, Narcissus Backdoor Attacks, BadNets). It then provides a concise introduction to the fundamentals of quantum optimization, including QUBO problems, the Ising model, the optimization principles of different quantum devices, and the implementation of neural networks trained using quantum computing.

\subsection{Attack Principles}

\subsubsection{Targeted Label-Flipping}

Targeted Label-Flipping attacks strategically alter the labels of training data, causing the model to learn incorrect label distributions during training and thereby affecting classification accuracy during testing.

\subsubsection{BadNets}

BadNets attacks involve inserting triggered samples into the training set. These samples have specific triggers (e.g., pixel patterns in images) added and are relabeled to a target class.

\subsubsection{Narcissus Backdoor Attacks}

Narcissus Backdoor Attacks insert samples into the training data that use ``salient regions" as triggers. This method combines salient features of input samples, making the backdoor trigger naturally integrated with the original sample features, thereby enhancing the stealthiness of the attack.

\subsection{QUBO Problems and Ising Model}

In the tutorial of ~\cite{glover2019tutorial} QUBO is a general mathematical modeling method that transforms complex optimization problems into a quadratic polynomial form:

\begin{equation}
\min \mathbf{x}^\top Q \mathbf{x} + \sum_{i} h_i x_i,
\end{equation}

where $\mathbf{x}$ is a binary decision variable vector, $Q$ is a symmetric weight matrix, and $h_i$ is the external penalty term for individual variables.

We consider an Ising system whose energy function can be expressed as:

\begin{equation}
E = \sum_{i > j} J_{ij} \sigma_i \sigma_j + \sum_{i} h_i \sigma_i,
\label{eq:ising_system}
\end{equation}

where $\sigma_i$ represents the spin state of the $i$-th qubit, taking values $\pm 1$. $J_{ij}$ is the coupling strength between spins, and $h_i$ is the external magnetic field. Generally, the minimum value of $E$ is sought, and the corresponding variable arrangement is determined in the process of minimization, that is, the value of $\sigma_i$. 
In the Ising model, the spin variables $\sigma_i \in \{-1, +1\}$ and the binary variables $x_i \in \{0, 1\}$ in the QUBO problem can be mapped through the following relationships:

\begin{equation}
x_i = \frac{1 + \sigma_i}{2}, \quad \sigma_i = 2x_i - 1.
\end{equation}

\subsection{Optimization of QUBO Problems Using Different Quantum Devices}

This subsection introduces three types of quantum devices capable of optimizing QUBO problems: Quantum Annealer, Coherent Quantum Photon Computer, and Gate Model Superconducting Quantum Computer.

\subsubsection{Quantum Annealer (QA)}

QAs by D-Wave ~\cite{Dwave1} are devices suitable for directly inputting coefficient matrices to solve QUBO problems. They simulate the quantum annealing process found in nature to search for global optimal solutions. Works by ~\cite{qa1,qa2} show that thanks to quantum tunneling effects, QAs can more easily find optimal solutions.

\subsubsection{Coherent Quantum Photon Computer (CQPC)}

CQPCs, especially Coherent Ising Machine (CIM) by QBoson Inc. ~\cite{kaiwu1,kaiwu2}, utilize the interference and entanglement characteristics of optical components to explore the solution space. Benefiting from optical design, CQPCS can perform searches from low to high energy states, thereby quickly finding the lowest energy state, which corresponds to the optimal solution ~\cite{cqpc1,cqpc2,cqpc3}.

\subsubsection{Gate Model Superconducting Quantum Computer}

Studies of ~\cite{QAOAtutorial,QAOA1,QAOA2} have shown that Gate Model Superconducting Quantum Computers are capable of running the Quantum Approximate Optimization Algorithm (QAOA). By transforming the QUBO problem into a quantum circuit and utilizing quantum gate operations to search for possible solutions, QAOA can be effectively executed on gate model quantum computers. Qiskit by IBM Inc. ~\cite{qiskit1} is one of QAOA simulators.

\subsection{Quantum Methods for Training Neural Network}
The famous term quantum machine learning, as shown in the study of ~\cite{qmltutorial1}, uses quantum circuit. Recently, several studies (\cite{energy1,energy2,energy3}) have shown that neural networks can be trained using energy-based methods. Among them, quantum computing devices can usually obtain a quantum solution by minimizing energy, and use the quantum solution to optimize the weight of the neural network to achieve the purpose of training the neural network. ~\cite{ep} has been using the ising machine and energy-based method to train neural networks.

\section{The Design of Q-Detection}

\begin{figure*}[htbp]
    \centering
    \includegraphics[scale=0.6]{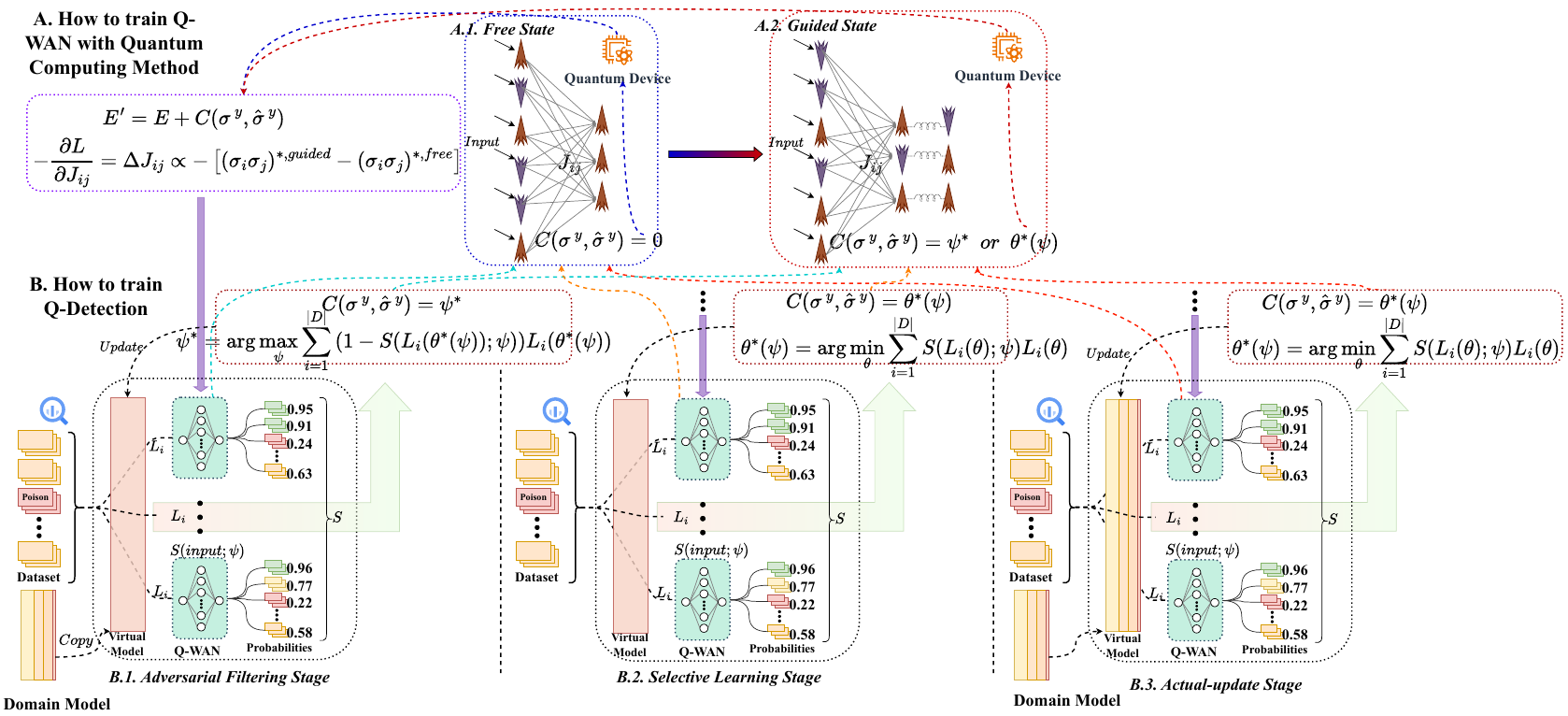}
    \caption{Q-Detection uses a one-hidden-layer network which is trained by quantum methods to filter poisoned data by assigning probabilities to samples. It maximizes the error for poisoned data in the Adversarial Filtering Stage and minimizes the error for clean data in the Selective Learning Stage and Actual-update Stage.}
    \label{fig:1}
\end{figure*}

This section introduces the design of Q-Detection, presenting the complete workflow and optimization methods in Fig.\ref{fig:1}. It also explains how the Q-WAN incorporates quantum computing through a bilevel QUBO optimization framework.

\subsection{Bilevel Optimization Process}

To formalize the above idea, we define the following optimization problem:

Suppose we have a poisoned dataset \( D = \{(x_i, y_i)\}_{i=1}^n \), where \( x_i \) represents the input sample and \( y_i \) is the corresponding label. We divide \( D \) into two subsets: \( B \) and \( D \setminus B \), where \( B \) is the clean dataset we aim to extract.

Since binary classification cannot be directly integrated into the network training process, we relax the binary variable \( v(k, i) \) indicating whether sample \( z_i \) belongs to training subset \( B \) to a continuous variable \( v(k, i) \in [0, 1] \), representing the probability that sample \( z_i \) is assigned as clean data. Here, \( k \) represents the training Stage.

The goal of Q-Detection is to train a Q-WAN \( S(\cdot; \psi) \) with parameters \( \psi \), which takes the model loss \( L_i(\theta) \) as input and outputs the weight for each sample. In this paper, the loss is the cross-entropy loss. This process can be formalized as the following bilevel optimization problem:

\begin{multline}
\psi^* = \arg\max_\psi \sum_{i=1}^{n} \left(1 - S(L_i(\theta^*(\psi)); \psi)\right) L_i(\theta^*(\psi)), \\
\text{subject to} \quad \theta^*(\psi) = \arg\min_\theta \sum_{i=1}^{n} S(L_i(\theta); \psi) L_i(\theta).
\label{eq:bilevel}
\end{multline}

where: \( \theta \)  represents the domain or its virtual model parameters and \( \psi \) represents the parameters of Q-WAN. \( L_i(\theta) \) denotes the loss of model \( \theta \) on sample \( z_i \). \( S(L_i(\theta); \psi) \) is the Q-WAN outputting the weight for sample \( z_i \).


Considering the difficulty of solving the constraint problem, we relax it as a two-step update process, including \textbf{Virtual-update} and \textbf{Actual-update} Stages. This is the process indicated by part B in Fig.\ref{fig:1}. The virtual-update Stage has two sub-stages including \textbf{Adversarial Filtering}, \textbf{Selective Learning} Stages.  The purpose of the virtual-update Stage is to give Q-WAN a training buffer so that it can assess whether incoming data is poisoned, preventing any direct harm to the domain model, and thus allowing Q-Detection to train properly.
Q-WAN will participate in every stage, learning how to assign weights to data by our QUBO model  and the training rules of Q-WAN.

Among them, the domain model replicates the virtual model, and the virtual model and Q-WAN jointly perform training in the Adversarial Filtering and Selective Learning Stage. The Actual-update Stage is jointly trained by the domain model and Q-WAN. This is because if Q-WAN is not well-trained, executing Eq.\ref{eq:bilevel} on the domain model will cause severe poisoning to the domain model, affecting the performance of Q-Dection.

\paragraph{Adversarial Filtering Stage}

In this Stage, we \textbf{maximize} the weighted error sum of data with a high probability of being poisoned (low-weight samples), penalizing the virtual model  from learning from poisoned data. The optimization objective is:

\begin{equation}
\arg\max_\psi \sum_{i=1}^{n} \left(1 - S(L_i(\theta^*(\psi)); \psi)\right) L_i(\theta^*(\psi))
\end{equation}

\paragraph{Selective Learning Stage}

In this Stage, we \textbf{minimize} the weighted error sum of data with a high probability of being clean (high-weight samples), encouraging the virtual model  to prioritize learning from clean data. The optimization objective is:

\begin{equation}
\arg\min_\theta \sum_{i=1}^{n} S(L_i(\theta); \psi) L_i(\theta)
\label{eq:selective}
\end{equation}

\paragraph{Actual-update Stage}

In this Stage, the same optimization of Selective Learning Stage 
 is taken, however encouraging the domain model  to prioritize learning from clean data. The optimization objective is similar to Eq. \ref{eq:selective}.

By alternating the optimization of \( \psi \) and \( \theta \), we gradually improve the model's adaptability to clean data while reducing dependence on poisoned data.

The above process is similar to Meta-Sift, but we introduce a Q-WAN, transforming the training of the weight-assigning network into a quantum optimization problem.

\subsection{QUBO Model of the Q-WAN}
Q-WAN training encodes the network as a QUBO, samples a “free” low-energy state and a “guided” state nudged toward target outputs on a quantum annealer, computes weight-wise correlation differences between the two states, and applies these differences as gradient updates in an iterative quantum-classical loop.

To transform the training process into a QUBO optimization problem, we redefine the training method of the weight-assigning network. By representing the training of the weight-assigning network as a QUBO problem, we can utilize quantum computing devices to solve it, achieving a quantum-classical hybrid poisoning detection method.

To incorporate \textbf{guided excitation} into the energy function of the Ising system, we modify the energy function to:

\begin{equation}
E' = E + C(\sigma^{\,y}, \hat{\sigma}^{\,y}),
\end{equation}

where:
\begin{itemize}
    \item \( E \) is as defined in Equation\ref{eq:ising_system}.
    \item \( y \) is the index set of output neurons.
    \item \( \hat{\sigma}_i \) is the target spin state (guided state).
    \item \( C(\sigma^{\,y}, \hat{\sigma}^{\,y}) \) represents the guided excitation term.
\end{itemize}

When the system is in the ground state (free evolution), \( C(\sigma^{\,y}, \hat{\sigma}^{\,y}) = 0 \); when guided excitation is introduced, \( C(\sigma^{\,y}, \hat{\sigma}^{\,y}) = \psi^* \) (Adversarial Filtering Stage) or \( C(\sigma^{\,y}, \hat{\sigma}^{\,y}) = \theta^*(\psi) \) (Selective Learning Stage and Actual-update Stage).

By introducing guided excitation into the energy function, we can minimize the spin energy during training, thereby optimizing the weight-assigning network. The introduction of guided excitation causes differences between the free state and guided state of the system, which can be used to compute gradients.

\subsection{Training Rules and Gradient Calculation}

To optimize the above objective, we design an innovative learning method. Using an equilibrium propagation-like approach, we derive the gradient direction from the discrete solution space by comparing the differences of the Q-WAN in the Free Stage and Guided Stage, and use it to update the parameters of the Q-WAN.

Here, we use the Mean Square Error (MSE) as the loss function, representing the deviation between the activation functions of the output neurons and their target states in the Q-WAN. We define two different measures \( L_1 \) and \( L_2 \) to evaluate the discrepancy between the activation states of output neurons and the target states, mathematically expressed as:

\begin{equation}
L_1(\rho(\mathbf{y}), \rho(\hat{\mathbf{y}})) = \arg\min_{\theta^*(\psi)} \|\rho(\mathbf{y}) - \rho(\hat{\mathbf{y}})\|^2,
\end{equation}

\begin{equation}
L_2(\rho(\mathbf{y}), \rho(\hat{\mathbf{y}})) = \arg\min_{-\psi^*} \|\rho(\mathbf{y}) - \rho(\hat{\mathbf{y}})\|^2,
\end{equation}

where:\( \mathbf{y} \) represents the set of output neurons. \( \rho(\mathbf{y}) \) denotes the activation function outputs of neurons.\( \hat{\mathbf{y}} \) is the target output state.

\paragraph{Measuring and Recording Spins}

We measure and record the spins at the end of the free and guided stages. The following learning rule is used to measure and record the changes in the states of output neurons and compute the gradient of the loss function with respect to the coupling strength \( J_{ij} \):

\begin{equation}
\label{eq10}
- \frac{\partial L}{\partial J_{ij}} = \Delta J_{ij} \propto - \left[ (\sigma_i \sigma_j)^{\text{guided}} - (\sigma_i \sigma_j)^{\text{free}} \right],
\end{equation}

where: \( (\sigma_i \sigma_j)^{\text{guided}} \) and \( (\sigma_i \sigma_j)^{\text{free}} \) represent the spin pairs in the guided state and free state, respectively.

We use these updates to adjust the parameters of the Q-WAN, allowing the network to gradually approach the optimal state. This is the process indicated by part A in Fig.\ref{fig:1}. The different states have solved by quantum devices and recorded the spins. The Q-WAN uses Eq.\ref{eq10} to update. 

\section{Experiment}

In this section, we design a series of experiments to verify the effectiveness of Q-Detection in defending against poisoning attacks and filtering clean data. The experiments are based on the GTSRB traffic sign image dataset, employing various common poisoning attack methods such as Targeted Label-Flipping attacks, BadNets attacks, and Narcissus backdoor attacks \cite{jha2023label,zeng2023narcissus,gu2019badnets} to evaluate the performance of Q-Detection.

The aim of the experiments is to, after poisoning the dataset, use defense methods like Q-Detection to filter out 4,000 clean data samples for training classification models (ResNet-18). 
Finally, the filtered datasets are used to train the same deep learning models, and the model performance is evaluated to test the quality of the filtered datasets.

\subsection{Experimental Setup}

\paragraph{Metrics:}

\textbf{Corruption Ratio (CR)} is a common metric used to measure the proportion of poisoned data in a dataset. The CR is calculated as:

\begin{equation}
\text{CR} = \frac{N_{\text{Poi}}}{|D_{\text{Sub}}|} \times 100\%,
\end{equation}

where \( N_{\text{Poi}} \) is the number of poisoned data points, and \( |D_{\text{Sub}}| \) denotes the size of the selected subset \( D_{\text{Sub}} \) from \( D \) using an automated method.

\textbf{Normalized Corruption Ratio (NCR)} is used as the evaluation metric to measure the ability to filter out poisoned data. NCR matrices normalize away the imbalance when detection performance is measured, allowing apples-to-apples comparisons across different attack methods. 
The NCR is calculated as:

\begin{equation}
\text{NCR} = \frac{\text{CR}}{\text{CR}_{\text{rand}}} \times 100\%,
\end{equation}

where \( \text{CR}_{\text{rand}} \) is the Corruption Ratio of a randomly selected subset, representing the capability of random filtering and serving as a random baseline.

A NCR value above \( 100\% \) indicates that the selection performance is worse than the naive random baseline. In contrast, a NCR value below \( 100\% \) implies better selection than random. The best possible NCR is \( 0\% \), indicating that the selected data is totally clean.

NCR value is used instead of attack success rate (ASR) to evaluate the quality of detection methods, because in most experiments, the SOTA method can achieve 0 poisoning detection. This means that the model will not be directly attacked.

\paragraph{General Settings}

All experiments were conducted on a computer with an Intel(R) Xeon(R) Gold 6226R CPU @ 2.90GHz with 64 GB RAM and an NVIDIA GeForce RTX 3090 24GB GPU. We utilized the Kaiwu SDK, D-Wave, and Qiskit Python API to solve the QUBO problem. Due to resource constraints, we mainly conducted  experiments in a simulated environment. The PyTorch framework and cuda was used for training classic models. 

We considered the popular benchmark dataset \textbf{GTSRB} (German Traffic Sign Recognition Benchmark) in ~\cite{dataset}. GTSRB is widely used for traffic sign recognition, containing images of 43 categories of traffic signs, with 39,209 training images and 12,630 test images.

We used \textbf{ResNet-18} as the base model for training on the GTSRB dataset. ResNet-18 is a lightweight deep convolutional neural network with good performance and generalization ability.

\subsection{How Clean the Selected Sub-dataset is}

\begin{table}[htbp]
\centering
\resizebox{\columnwidth}{!}{%
\begin{tabular}{c|c|ccccccc}
\toprule
\textbf{Attack Type} &
\textbf{\makecell{Poisoning \\ Ratio (\%)}} &
\textbf{DCM} &
\textbf{\makecell{Loss \\ Scan}} &
\textbf{\makecell{AutoEncoder \\ Outlier}} &
\textbf{\makecell{Meta-\\Sift}} &
\textbf{\makecell{Q-Det. CQPC \\ 500 qubits}} &
\textbf{\makecell{Q-Det. QA \\ 500 qubits}} &
\textbf{\makecell{Q-Det. QA \\ 5000 qubits}} \\
\midrule
\multirow{5}{*}{\makecell{Targeted \\ Label-\\ Flipping}}
& 3  & 15.24\% & 0.00\% & 76.58\% & \textbf{0.00\%} & 0.00\% & 0.00\% & \textbf{0.00\%} \\
& 5  & 18.06\% & 0.00\% & 108.91\% & \textbf{0.00\%} & 0.00\% & 0.00\% & \textbf{0.00\%} \\
& 10 & 38.17\% & 0.00\% & 106.55\% & \textbf{0.00\%} & 0.00\% & 0.00\% & \textbf{0.00\%} \\
& 20 & 62.27\% & 5.61\% & 115.65\% & \textbf{0.00\%} & 0.00\% & 0.00\% & \textbf{0.00\%} \\
& 30 & 87.99\% & 6.59\% & 118.25\% & \textbf{0.00\%} & 0.00\% & 0.00\% & \textbf{0.00\%} \\
\midrule
\multirow{5}{*}{\makecell{Narcissus \\ Backdoor}}
& 3  & 0.00\% & 94.39\% & 0.00\% & \textbf{0.00\%} & 0.00\% & 0.00\% & \textbf{0.00\%} \\
& 5  & 0.00\% & 132.57\% & 0.00\% & \textbf{0.00\%} & 0.00\% & 0.00\% & \textbf{0.00\%} \\
& 10 & 0.00\% & 235.59\% & 0.00\% & \textbf{0.00\%} & 0.00\% & 0.00\% & \textbf{0.00\%} \\
& 20 & 0.00\% & 197.89\% & 0.00\% & \textbf{0.00\%} & 0.00\% & 0.00\% & \textbf{0.00\%} \\
& 30 & 0.00\% & 188.47\% & 0.00\% & \textbf{0.00\%} & 180.00\% & 267.00\% & \textbf{0.00\%} \\
\midrule
\multirow{6}{*}{\makecell{BadNets \\ Backdoor}}
& 1  & 0.00\% & 0.00\% & 0.00\% & \textbf{0.00\%} & 0.00\% & 0.00\% & \textbf{0.00\%} \\
& 3  & 30.49\% & 0.00\% & 0.00\% & \textbf{0.00\%} & 0.00\% & 0.00\% & \textbf{0.00\%} \\
& 5  & 18.05\% & 0.00\% & 0.00\% & \textbf{0.00\%} & 0.00\% & 0.00\% & \textbf{0.00\%} \\
& 8  & 16.24\% & 0.00\% & 0.00\% & \textbf{0.00\%} & 0.00\% & 0.00\% & \textbf{0.00\%} \\
& 10 & 16.95\% & 0.00\% & 0.00\% & \textbf{0.00\%} & 110.14\% & 330.00\% & \textbf{0.00\%} \\
& 30 & 57.21\% & 121.07\% & 0.00\% & \textbf{0.00\%} & 231.24\% & 365.57\% & \textbf{0.00\%} \\
\bottomrule
\end{tabular}}
\caption{Q-Det. is the abbreviation of Q-Detection in this table. 
We compare several defence methods against three types of poisoning attacks while varying the available qubits (500 and 5000).
Each Q-WAN uses the same number of hidden-layer neurons as qubits and is trained with simulated quantum computing sdk.
Because CQPC simulated SDK currently supports no more than 600 qubits, only the 500-qubit setting is reported for that solver.
At 500 qubits, both Q-Det. CQPC and Q-Det. QA fall behind the SOTA method on backdoor attacks with higher poisoning ratios.
 In contrast, Q-Det. with 5000 hidden layer neurons trained on 5000 qubits performs significantly better under higher levels of attack and achieves SOTA at 0\% NCR, illustrating how Q-Detection gains from advances in quantum-computing hardware.}
\label{table:results}
\end{table}

For the GTSRB dataset, we applied Targeted Label-Flipping attacks, BadNets attacks, and Narcissus backdoor attacks to poison the data. Then, we used various filtering methods to remove poisoned data, comparing the NCR metrics of Q-Detection with other methods such as DCM and Meta-Sift.

We filtered out 4,000 data samples as the clean dataset.

In Q-Detection, we set 20, 500, and 5000 neurons as the hidden layer of the Q-WAN, corresponding to the number of qubits in the simulation. We used local devices and the Kaiwu SDK (600 qubits limitation), D-Wave SDK (5000+ qubits limitation), and Qiskit (30 qubits limitation) for solving QUBO. Packages that exceed the qubits limit were not used.

After training, we obtained the trained Q-WAN and used it to filter clean data. Comparing with other filtering methods, we obtained the results shown in Table\ref{table:results}. 
It is worth noting that in the experiment with 20 neurons, the three training methods all failed. Qiskit's QAOA method could only train 20 neurons and obtained a 929\% NCR, which is due to the small size of the hidden layer and the insufficient representation ability of the neural network. Additionally, ~\cite{chu2023qtrojan,li2025qpoison} showing that quantum circuits are inherently vulnerable to circuit poisoning attacks, we propose using Quantum Annealer or CQPC with more qubits.

The tables show the NCR metrics of different defense methods under three representative poisoning attacks. The SOTA method Meta-Sift achieves the best NCR metrics under all attack types, demonstrating significant advantages in filtering clean data.

Q-Detection maintains an advantage over other baseline methods and achieves SOTA with 5000 qubits. Based on the success of simulated quantum computing, it shows potential for faster solutions on real quantum hardware in the future.

\begin{figure*}[htbp]
    \centering
    \includegraphics[width=0.9\textwidth]{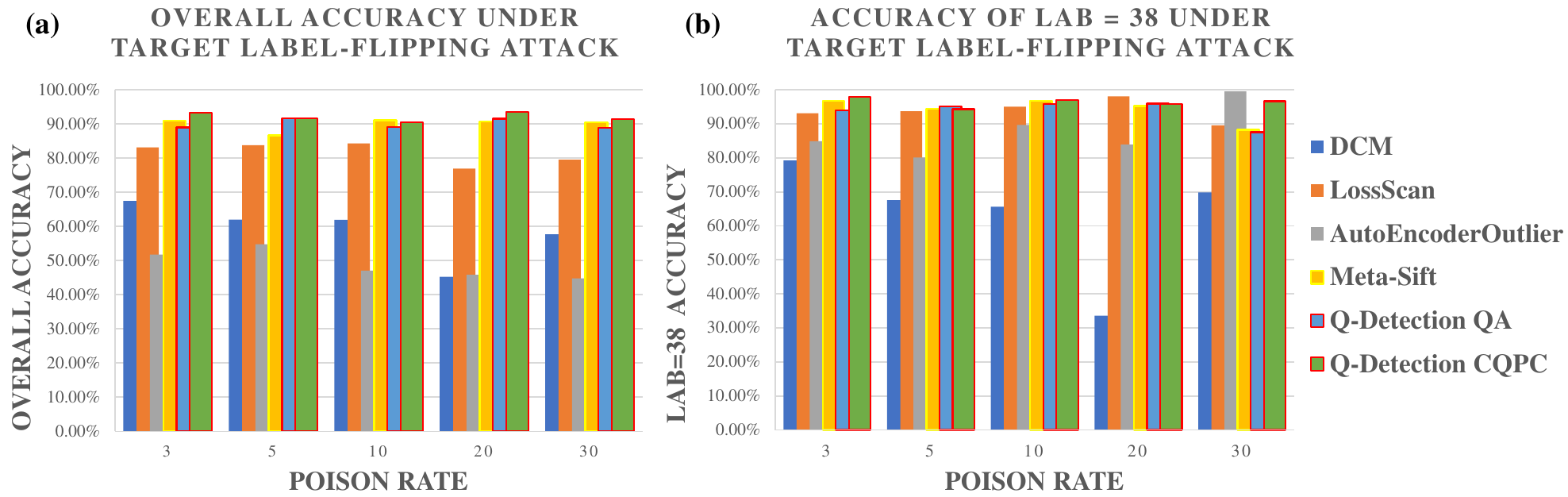}
    \caption{Under the Target Label flipping attack, we consider the poisoning ratio from 3\% to 30\%, and Q-WAN has 500 hidden layer neurons, which consumes 500 qubits. 4000 clean data are extracted using the trained Q-WAN to form the clean data subset, and then a new ResNet-18 is trained using the clean data subset to test its accuracy of overall and Target label equal to 38.}
    \label{fig:acc-result}
\end{figure*}

Other baseline methods have various drawbacks. The DCM method is only effective against Narcissus Backdoor attacks and performs poorly against other attacks. The LossScan method is extremely ineffective in filtering Narcissus Backdoor attacks. The Autoencoder Outlier method performs poorly under BadNets Backdoor attacks and takes about 30 minutes to train, which is longer than Q-Detection and Meta-Sift.

\subsection{Train a New Model on the Selected Sub-Dataset}

After filtering, the selected sub-dataset is used for training a brand new deep learning model and testing its accuracy. Higher model accuracy demonstrates high-quality of sub-dataset. 

The experimental results in Fig.\ref{fig:acc-result} show that Q-Detection keeps ahead of Meta-Sift in overall accuracy, achieving the SOTA effect. In the accuracy test of label is 38, Q-Detection trained with CQPC still keeps the lead over Meta-Sift. Q-Detection is superior to SOTA method most of the time, for example, when poison rate is 20\% and 30\%, Compared with the model trained on the dataset screened by Meta-Sift, the overall accuracy of Q-Detection CQPC is improved by about \textbf{3\%} and \textbf{1\%} in Fig.\ref{fig:acc-result} part (a).

\subsection{Quantitative Estimate of the Time Superiority}

Since Q-Detection requires multiple calls to quantum computing devices, and current quantum computing devices are not well-integrated with GPUs, we perform a theoretical estimation of Q-Detection under the Targeted Label-Flipping Attack to speculate on the potential acceleration that quantum computing can bring to the field. We run Q-Detection and Meta-Sift multiple times, take their average values, and use the average simulated computing time as a reference.

We conducted tests on QBoson Inc's 550-qubit CQPC, obtaining an average time difference between simulated quantum computing and real machines of \(10^3\). 
Similarly, on D-Wave's Quantum Annealer, the real machine computation time is significantly less than that of simulated quantum computing ~\cite{qrealtime1,qrealtime2,kaiwu1}.




We set the batch size to 179, requiring each Q-WAN to train for 220 epochs. \textbf{Notably}, using simulated QA can achieve an average training speed of 1 minute and 10 seconds, which is already very close to the CUDA-accelerated Meta-Sift training time. This suggests promising direct acceleration effects on real Quantum Annealer hardware.

Reducing the batch size lowers memory requirements, necessitating more training epochs and increasing the number of calls to quantum computing devices. 

When we set the batch size to approximately one-quarter of the original size, i.e., 44, each Weight-Assigning Network needs to train for 892 epochs, resulting in an overall training time of 69 minutes. In contrast, Meta-Sift's average training duration is 3 minutes. Each batch of data training still consumes 4.5 seconds of computation time, requiring a total of 3,568 calls to quantum computing devices and spending 66.9 minutes on simulated quantum computing. Thus, we estimate the total training time with integrated CIM computing power to be approximately \(2.1 + 0.0669 = 2.17\) minutes, achieving over a 20\% acceleration compared to Meta-Sift purely accelerated by CUDA.


\section{Related Work}
To address the impact of data poisoning on image classification models, researchers have proposed various defense methods, especially in the tasks of clean data selection and anomaly detection.

Some typical methods attempt to distinguish poisoned samples from clean samples by analyzing sample features, loss patterns, or data distribution characteristics. The following are several representative research methods:

\begin{enumerate}
    \item \textbf{DCM (Distance to Class Means)}
    
\textbf{Method Overview:} The DCM method of ~\cite{dcm1} is a clustering-based approach. It extracts the output vectors of samples from the last hidden layer of a proxy model as feature representations. By calculating the distance between each sample and the mean of its class, clean samples are filtered out. It is a simple and effective method, which we consider as a baseline method.

Specifically, the mean sample (class center) of each category is first calculated, and then the distance from each sample to the class center is computed. Samples with smaller distances are regarded as clean samples.

\textbf{Advantages:} Simple and efficient, particularly suitable for datasets with regular feature distributions.

\textbf{Disadvantages:} May perform poorly in scenarios with imbalanced class distributions or severely poisoned data.

\item \textbf{LossScan (Poisoning Scan Based on Feature Loss Sensitivity)}

\textbf{Method Overview:} LossScan of ~\cite{LossScan} utilizes the loss characteristics of samples in the early stages of training to determine whether they are poisoned. Studies have shown that poisoned samples often exhibit abnormal losses during early training. By recording training losses and filtering out samples with abnormally high losses, poisoned samples can be effectively removed.

\textbf{Advantages:} Simple to implement, no need for complex modeling, can efficiently utilize information generated during the training process.

\textbf{Disadvantages:} Relies on the training process, increasing training time.

\item \textbf{Autoencoder-based Outlier Detection}

\textbf{Method Overview:} The autoencoder-based outlier detection method by ~\cite{autoencoder} constructs an unsupervised model to learn low-dimensional representations of the data and uses reconstruction error to identify anomalous samples. Anomalous samples typically have higher reconstruction errors and can thus be filtered out as potential poisoned samples.

\textbf{Advantages:} It is suitable for unsupervised scenarios, and has strong adaptability.

\textbf{Disadvantages:} Autoencoders may overfit to anomalous samples, affecting detection performance.

\item \textbf{Meta-Sift}

\textbf{Method Overview:} Meta-Sift by ~\cite{metasift} innovatively proposes a bilevel optimization framework to extract clean subsets from poisoned datasets. The inner optimization objective is to minimize the training loss of the selected subset, while the outer optimization objective is to maximize the prediction loss on the unselected data. Through this mechanism, poisoned samples can be effectively identified, enhancing the robustness of the classification model.

Meta-Sift is a novel and efficient method, which we consider as the state-of-the-art (SOTA) method.

\textbf{Advantages:} Does not rely on clean benchmark data, can provide efficient defense.

\textbf{Disadvantages:} The computational complexity of bilevel optimization is high, suitable for scenarios with abundant computational resources.
\end{enumerate}

The above methods provide different perspectives and technical solutions for filtering clean datasets. Existing methods rely on classical computing frameworks and may be limited by computational resources and efficiency when handling high-dimensional data and complex scenarios. Many works( ~\cite{quantumspeed1,quantumspeed2,quantumspeed3} ) in the field of quantum computing have demonstrated the potential of quantum computing to accelerate the solution of various problems.

\section{Conclusion}

Through simulated experiments and theoretical estimations, we have verified the acceleration that integrating quantum computing can bring to the field of data poisoning detection. Furthermore, we reiterate that the Q-Detection method benefits from the development of quantum computing technology, as evidenced by the enhanced representation capability of the Q-WAN with more qubits and improved compatibility of the quantum-classical hybrid architecture. In the future, real machine experiments are expected to bring actual acceleration capabilities.

\section*{Acknowledgments}
This work was supported by Shenzhen Science and Technology Program (No.KJZD20240903095700001), National Natural Science Foundation of China (No.62441619), and the Fundamental Research Funds for the Central Universities, Sun Yat-sen University under Grants No.23xkjc010.

\bibliographystyle{named}
\bibliography{ijcai25}

\end{document}